\documentclass[aps,prd,twocolumn,superscriptaddress,nofootinbib]{revtex4-1}

\usepackage{latexsym}
\usepackage{amsmath}
\usepackage{amssymb}
\usepackage{amsfonts}
\usepackage{bm}

\usepackage{color}
\definecolor{purple}{rgb}{0.5,0,0.5}
\definecolor{blue}{rgb}{0.0,0,0.9}
\definecolor{prdblue}{rgb}{0.133,0.118,0.498}
\usepackage[colorlinks=true, pdfstartview=FitV, linkcolor=prdblue, citecolor= prdblue, urlcolor=prdblue]{hyperref}

\usepackage{supertabular} 
\usepackage{placeins}
\usepackage{epsfig}
\usepackage{graphicx}

\begin{document}


\title{Exploring doubly-heavy tetraquarks in constituent-quark-model based \\ meson-meson coupled-channels approach}


\author{P. G. Ortega}
\email[]{pgortega@usal.es}
\affiliation{Departamento de F\'isica Fundamental, Universidad de Salamanca, E-37008 Salamanca, Spain}
\affiliation{Instituto Universitario de F\'isica 
Fundamental y Matem\'aticas (IUFFyM), Universidad de Salamanca, E-37008 Salamanca, Spain}

\author{D. R. Entem}
\email[]{entem@usal.es}
\affiliation{Departamento de F\'isica Fundamental, Universidad de Salamanca, E-37008 Salamanca, Spain}
\affiliation{Instituto Universitario de F\'isica Fundamental y Matem\'aticas (IUFFyM), Universidad de Salamanca, E-37008 Salamanca, Spain}

\author{F. Fern\'andez}
\email[]{fdz@usal.es}
\affiliation{Instituto Universitario de F\'isica Fundamental y Matem\'aticas (IUFFyM), Universidad de Salamanca, E-37008 Salamanca, Spain}

\author{J. Segovia}
\email[]{jsegovia@upo.es}
\affiliation{Departamento de Sistemas Físicos, Químicos y Naturales, Universidad Pablo de Olavide, E-41013 Sevilla, Spain}

\date{\today}

\begin{abstract}
The LHCb Collaboration announced in 2021 the discovery of a new tetraquark-like state, named $T_{cc}^+$, with minimum quark content $cc\bar u\bar d$, close to the $D^0D^{*+}$ threshold. This has motivated countless theoretical works trying to identify the dynamics which is responsible of the formation of such state; in particular, the one performed by us in Ref.~\cite{Ortega:2022efc}, where a $D^0D^{*+}$ molecular candidate whose mass, width, scattering length and effective ranges are in reasonable agreement with experimental measurements.
We explore herein the possibility of having $T_{cc}^+$ partners in all doubly-heavy tetraquark sectors, considering doubly represented light antiquarks $u$, $d$ or $s$, and taking into account all possible spin-parity quantum numbers.
The computation is done using a constituent-quark-model based meson-meson coupled-channels framework which has been tested many times in the last fifteen years describing conventional heavy mesons and baryons, their coupling with hadron-hadron thresholds but also in exploring its application to compact multiquark structures.
The advantage of using an approach with such a relatively large history is that it allows us to make predictions because all the parameters have already been constrained from our previous works. Then, from this perspective, we present a parameter-free model-dependent prediction of doubly-heavy tetraquarks that may be partners of the discovered $T_{cc}^+$ state.
\end{abstract}


\maketitle


\section{INTRODUCTION}
\label{sec:introduction}

Prior to the turning of the XXI century, the spectrum of identified hadrons was restricted to systems that correspond to the patterns typically described by constituent-quark models~\cite{Gell-Mann:1964ewy, Zweig:1964CERN}; that is to say, hadrons were basically separated into mesons and baryons which are, respectively, colorless systems of quark-antiquark and three-quarks. Notwithstanding this, the authors of Refs.~\cite{Gell-Mann:1964ewy, Zweig:1964CERN} also raised the possibility of having more complicated arrangements of quarks inside hadrons such as tetraquarks, pentaquarks and even hexaquarks. Since then, investigations on the phenomenology of multiquark systems were conducted; particularly relevant was the decade of the 1970s when Jaffe, among others, published a coupled of seminal works~\cite{Jaffe:1976ig, Jaffe:1976ih} proposing, for instance, a four quark interpretation for the lightest scalar meson.

A large amount of data taken in the last twenty years by B-factories (BaBar, Belle and CLEO), $\tau$-charm facilities (CLEO-c and BESIII) and also proton--(anti-)proton colliders (CDF, D0, LHCb, ATLAS and CMS), has provided evidence for the possible existence of dozens of exotic charmonium- and bottomonium-like states, the so-called XYZ mesons. In addition, these prominent experimental programmes have stimulated extensive theoretical investigations in the field as well as the breadth and depth necessary for a vibrant research environment in contemporary particle physics. For a comprehensive description of the current status of heavy quarkonium physics, the reader is referred to several reviews~\cite{Lebed:2016hpi, Chen:2016qju, Chen:2016spr, Ali:2017jda, Guo:2017jvc, Olsen:2017bmm, Liu:2019zoy, Brambilla:2019esw, Yang:2020atz, Dong:2020hxe, Dong:2021bvy, Chen:2021erj, Cao:2023rhu, Mai:2022eur, Meng:2022ozq, Chen:2022asf, Guo:2022kdi, Ortega:2020tng, Huang:2023jec, Lebed:2023vnd, Zou:2021sha}.
 
Among the mentioned exotic hadrons, those with manifestly exotic heavy-flavor content, having four quarks in their valence structure, are extremely interesting. Several of them are in fact close to open-flavor meson-meson thresholds, suggesting to a connection with the corresponding scattering channel for their existence, and possibly their nature. In particular, we need to mention the 2021 announcement of the LHCb collaboration of a new tetraquark-like state, named $T_{cc}^+$, with minimum quark content $cc\bar u\bar d$, close to the $D^{*+}D^0$ threshold~\cite{LHCb:2021vvq}. Its importance relies on its doubly charmed content, in contrast with the hidden-charm nature of all previously observed exotic states such as the first observed XYZ-state, \emph{viz.} the $X(3872)$~\cite{Belle:2003nnu}. The mass and width of $T_{cc}^+$, obtained from a Breit-Wigner parametrization, is
\begin{subequations}
\begin{align}
\delta m_{\rm BW} &= (-273\pm61\pm5^{+11}_{-14})\, {\rm keV/c}^2 \,, \label{eq:BWexp1}\\
\Gamma_{\rm BW} &= (410\pm165\pm43^{+18}_{-38})\,{\rm keV} \,. \label{eq:BWexp2}
\end{align}
\end{subequations}
where $\delta m_{\rm BW}=m_{T_{cc}^+}-m_{D^{*+}D^0}$. In a further analysis~\cite{LHCb:2021auc}, an estimation of the pole position in the complex energy plane provides $\delta m_{\rm pole}=(-360\pm40^{+4}_{-0})$~keV/c$^2$ and $\Gamma_{\rm pole}=(48\pm2^{+0}_{-14})$~keV. That is to say, the mass is compatible with the previous measurement of Eq.~\eqref{eq:BWexp1} but the width is much lower, see Eq.~\eqref{eq:BWexp2}, which is reasonable due to the limited phase space of the  $D^0D^0\pi^+$ decay. This may indicate that the $\sim\!\!400$ keV width of Ref.~\cite{LHCb:2021vvq} is a consequence of the resolution function, whose root mean square is actually around $400$ keV~\cite{LHCb:2021vvq, LHCb:2021auc}. Thus, we have an extremely narrow state, very close to threshold, which is a strong candidate for a pure hadron molecule.

It is fair to mention that the $T_{cc}^+$ state was already predicted in the 1980s and 1990s (see, \emph{e.g.}, Refs.~\cite{Ader:1981db, Zouzou:1986qh, Lipkin:1986dw, Heller:1986bt, Manohar:1992nd}). Its recent discovery has also inspired a large amount of new theoretical studies~\cite{Meng:2022ozq, Maiani:2022qze, Lin:2022wmj, Cheng:2022qcm, Achasov:2022onn, Padmanath:2022cvl, Agaev:2022ast, Yang:2019itm, Deng:2021gnb, Santowsky:2021bhy, Du:2021zzh, Baru:2021ldu, Albaladejo:2021vln, Ren:2021dsi, Xin:2021wcr, Guo:2021yws, Ling:2021bir, Agaev:2021vur, Qin:2020zlg, Meng:2023jqk, Whyte:2024ihh} that associate such state to either a $DD^*$ molecule or a compact tetraquark (see the review of Ref.~\cite{Chen:2022asf}, and references therein, for further details). Nevertheless, the proximity of the $DD^\ast$ threshold could mean that, although there may exist a compact tetraquark core, the dominant component be the molecular one, so the description of this state as a pure $DD^\ast$ molecule is more than justified.

In Ref.~\cite{Ortega:2022efc}, we performed a coupled-channels calculation of the $J^P=1^+$ $cc\bar q\bar q'$ sector and find a molecular candidate for the $T_{cc}^+$ whose mass, with respect to the lower $D^0D^{\ast+}$ threshold, and width are, respectively, $-387\,\text{keV}$ and $81\,\text{keV}$. The molecule is a about $87\%$ $D^0D^{\ast+}$ system with a sizable isospin breaking due to the mass difference of the $D^0D^{\ast+}$ and $D^+D^{\ast0}$ thresholds. We also analysed the scattering length and effective ranges of the molecule and found them to be in agreement with the LHCb analysis. We then want to expand our analysis to the doubly-heavy tetraquarks, \emph{i.e.} $QQ'\bar q\bar q$ with $Q,Q'=\{c,b\}$ and $q=\{u,d,s\}$.

The coupled-channels calculation is based on a constituent quark model (CQM)~\cite{Vijande:2004he, Segovia:2008zz} which has been widely used in heavy quark sectors, studying their spectra~\cite{Segovia:2011zza, Segovia:2015dia, Yang:2019lsg, Ortega:2020uvc}, their electromagnetic, weak and strong decays and reactions~\cite{Segovia:2011zza, Segovia:2011dg, Segovia:2014mca, Martin-Gonzalez:2022qwd}, and meson-meson, baryon-meson and baryon-baryon states and the coupling of meson-meson states with the meson spectrum~\cite{Ortega:2012rs,Ortega:2016pgg, Ortega:2018cnm, Ortega:2021xst, Ortega:2021fem}, see also the reviews~\cite{Segovia:2013wma, Ortega:2020tng}. The advantage of using a model with relatively large history is that it allows us to make predictions. Indeed, all the parameters of the model have already been constrained from previous works. Then, from this perspective, we present a parameter-free model-dependent prediction of doubly-heavy tetraquarks that may be partners of the discovered $T_{cc}^+$ state.

The manuscript is structured as follows. After this introduction, the theoretical framework is briefly presented in section~\ref{sec:theory}. In section~\ref{sec:results}, we analyze and discuss our theoretical results. Finally, we summarize and draw some conclusions in Sec.~\ref{sec:summary}.


\section{THEORETICAL FORMALISM}
\label{sec:theory} 

\subsection{Constituent quark model}

For the study of $QQ^\prime \bar q\bar q$ systems with $Q,Q'=\{c,b\}$ and $q=\{n,s\}$, where $n=\{u,d\}$, we employ the same theoretical framework as in Ref.~\cite{Ortega:2022efc}. This is to say, a constituent quark model (CQM) whose main features are the dynamical chiral symmetry breaking of the Lagrangian of Quantum Chromodynamics (QCD), the perturbative one-gluon exchange (OGE) contribution and the non-perturbative confinement of quarks. Detailed expressions of the potentials and the model parameters of the CQM can be found in, for instance, Refs.~\cite{Vijande:2004he, Segovia:2008zz}.

For convenience, we briefly describe here the most relevant characteristics of the aforementioned model, relevant for the description of meson-meson systems. The core of the CQM is rooted in the development of a dynamical quark mass resulting from the spontaneous breaking of the chiral symmetry in QCD. According to the Goldstone theorem, this leads to the emergence of boson exchanges among light quarks, a phenomenon that can be effectively represented using the chiral invariant effective Lagrangian~\cite{Diakonov:2002fq}
\begin{equation}
{\mathcal L} = \bar{\psi}(i\, {\slash\!\!\! \partial}
-M(q^{2})U^{\gamma_{5}})\,\psi  \,.
\end{equation}
The momentum-dependent dynamical quark mass is shown as $M(q^2)$, which tends to $\sim\! 350$ MeV for light $u$ and $d$ quarks for $q\to 0$. The Goldstone-boson fields, $U^{\gamma_5}= e^{i\lambda _{a}\phi ^{a}\gamma _{5}/f_{\pi}}$, with $\phi=\{\vec \pi,K,\eta_8\}$, appear to preserve chiral invariance of the QCD Lagrangian and describe the interaction between constituent light quarks. Note, however, that for heavy-heavy ($QQ$) or heavy-light ($Qq$) quark pairs, Goldstone-boson exchanges are not allowed as chiral symmetry is explicitly broken. 

Besides the non-pertubative interaction due to Goldstone bosons, the model incorporates a one-gluon exchange potential, extracted from the vertex Lagrangian
\begin{equation}
{\mathcal L}_{qqg} = i\sqrt{4\pi\alpha_{s}} \, \bar{\psi} \gamma_{\mu}
G^{\mu}_{c} \lambda^{c} \psi \,,
\label{Lqqg}
\end{equation}
with $\lambda^{c}$ the $SU(3)$ color matrices and $G^{\mu}_{c}$ the
gluon field. The $\alpha_{s}$ is a scale-dependent effective strong coupling constant that allows a comprehensive description of light, strange
and heavy meson spectra (see, \emph{e.g.}, Ref.~\cite{Segovia:2008zz} for its explicit implementation).

Finally, a screened confinement term is incorporated to ensure colorless hadrons. The potential is linearly-rising for short interquark distances, but acquires a plateau at large distances to mimic the effect of sea quarks, which induces the breakdown of the $q\bar q$ color binding string~\cite{Bali:2005fu}. Its specific expression is
\begin{equation}
V_{\rm CON}(\vec{r}\,)=\left[-a_{c}(1-e^{-\mu_{c}r})+\Delta \right]
(\vec{\lambda}_{q}^{c}\cdot\vec{\lambda}_{q}^{c}) \,,
\label{eq:conf}
\end{equation}
being $a_{c}$, $\Delta$ and $\mu_{c}$ model parameters.
Above the screened potential plateau, the meson undergoes a transition from a $q\bar q$ colorless configuration into a pair of mesons due to the breaking of the color string.

All model parameters are constrained based on prior investigations into hadron phenomenology~\cite{Vijande:2004he, Segovia:2008zz}, rendering the calculation presented herein effectively parameter-free. This enables us to provide reliable predictions regarding the presence or absence of particular molecular configurations. These parameters were fixed to describe a set of hadron observables within a certain range of agreement and thus a theoretical uncertainty is associated with this model adjustment. To assess this, the results presented in this manuscript show a theoretical uncertainty due to a variation of $\pm 10\%$ in the strength of the CQM potentials.

\subsection{Resonating group method}

The interaction between quarks and (anti-)quarks described by the CQM allows us to obtain the masses and wave functions of the $D_{(s)}^{(*)}$ and $\bar B_{(s)}^{(*)}$ mesons when solving the Schr\"odinger equation with the CQM's potentials using the Gaussian Expansion Method~\cite{Hiyama:2003cu}. To characterize the interaction between two $H_Q$ $(\equiv Q\bar q)$ mesons, we use the Resonating Group Method (RGM)~\cite{Wheeler:1937zza, Tang:1978zz}. In this approach, the meson-meson molecular interaction arises as a residual effect of the underlying quark dynamics (see, \emph{e.g.}, Refs.~\cite{Fernandez:2019ses} for a detailed explanation).

\begin{figure}[!t]
\hspace*{-.5cm}\includegraphics[width=.5\textwidth]{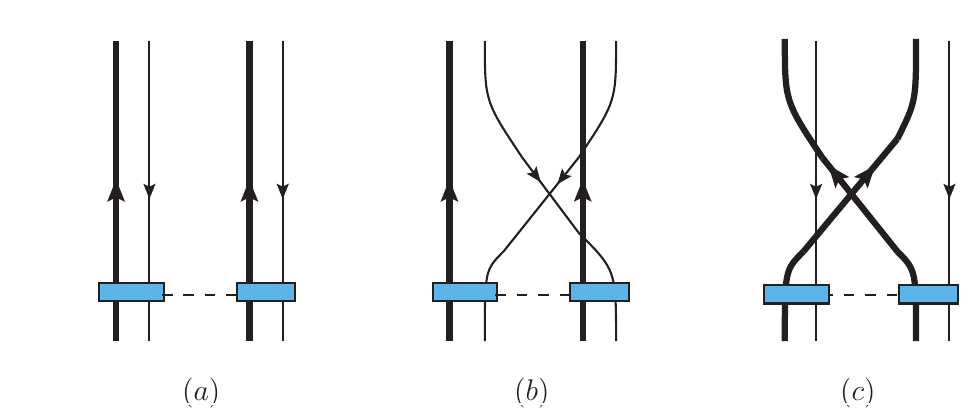}
\caption{\label{fig:diagrams} Type of diagrams considered in this work:  \emph{Panel (a)}: Direct diagrams (Eq.~\eqref{eq:directV}), \emph{Panel (b)}: Exchange of the two light-quark pairs (Eq.~\eqref{eq:exchangeV}) and, \emph{Panel (c)}: Exchange of the two heavy-quark pairs. Heavy (light) quarks are shown as thick (thin) lines. The blue bands represent the sum of interactions between constituents of different clusters.}
\end{figure}

All considered $H_QH_{Q^\prime}$ systems in this work have a pair of identical light or strange quarks, while some of them also have identical heavy quarks, so we must ensure that the full wave function is totally antisymmetric. Thus, the total wave function of a $(H_QH_{Q^\prime})$-system should be built as
\begin{align}
\Psi = {\cal A}\left[\phi_{A}\phi_{B}\chi_L\sigma_{ST}\xi_c\right] \,,
\end{align}
where ${\cal A}$ is the fully antisymmetric operator, $\phi_{A(B)}$ is the wave functions of the $A(B)$ meson, $\chi_L$ the relative orbital wave function of the $AB$ pair, $\sigma_{ST}$ their spin-isospin wave function and $\xi_c$ their color wave function. Besides, the $H_cH_b$ fully antisymmetric operator can be written, up to a normalization factor, as ${\cal A}=(1-P_q)$, where $P_q$ is the operator that exchanges light antiquarks (Fig.~\ref{fig:diagrams}(b)), whereas for the $H_cH_c$ or $H_bH_b$ systems the operator is ${\cal A}=(1-P_q)(1-P_Q)$~\cite{Ortega:2022efc}.

Hence, the meson-meson residual interaction can be factorized in a direct potential, without quark rearrangements between clusters, and an exchange kernel, where such reordering occurs. Then, the process $AB\to A'B'$ can be modeled as
\begin{align}\label{eq:fullpot}
 ^{\rm RGM}V(\vec P',\vec P) = \,^{\rm RGM}V_D(\vec P',\vec P) + \,^{\rm RGM}K(\vec P',\vec P) \,,
\end{align}
where $\vec P^{(\prime)}$ is the initial (final) relative momentum of the $AB$ ($A'B'$) system, $^{\rm RGM}V_D$ is the direct potential and $^{\rm RGM}K$ is the exchange kernel.

The direct kernel (Fig.~\ref{fig:diagrams}(a)) can be obtained as,
\begin{align}\label{eq:directV}
&
{}^{\rm RGM}V_{D}(\vec{P}',\vec{P}) = \sum_{i\in A, j\in B} \int d\vec{p}_{A'} d\vec{p}_{B'} d\vec{p}_{A} d\vec{p}_{B} \times \nonumber \\
&
\times \phi_{A'}^{\ast}(\vec{p}_{A'}) \phi_{B'}^{\ast}(\vec{p}_{B'})
V_{ij}(\vec{P}',\vec{P}) \phi_A(\vec{p}_{A}) \phi_B(\vec{p}_{B})  \,,
\end{align}
with $V_{ij}$ the CQM quark-quark interaction, $\vec p_{A(B)}$ the relative internal momentum of the $A$ ($B$) meson and $i$ $(j)$ the indices of the constituents of the $A$ ($B$) meson. One may appreciate that there is a natural cut-off for the potential given by the wave functions of the involved mesons.

The quark rearrangement between the two clusters is modeled by a non-local energy-dependent exchange kernel $^{\rm RGM}K$, given by
\begin{align}
{}^{\rm RGM}K(\vec P',\vec P_i) &= {}^{\rm RGM}H_E(\vec P',\vec P_i) \nonumber \\
&
- E_T\,{}^{\rm RGM} N_E(\vec P',\vec P_i)\,,
\end{align}
which is split in a potential plus a normalization term, where $E_T$ is the total energy of the system and $\vec P_i$ is a parameter. The exchange Hamiltonian and normalization terms can be written as
\begin{subequations}
\begin{align}\label{eq:exchangeV}
&
{}^{\rm RGM}H_{E}(\vec{P}',\vec{P}_{i}) = \int d\vec{p}_{A'}
d\vec{p}_{B'} d\vec{p}_{A} d\vec{p}_{B} d\vec{P} \phi_{A'}^{\ast}(\vec{p}_{A'}) \times \nonumber \\
&
\times  \phi_{B'}^{\ast}(\vec{p}_{B'})
{\cal H}(\vec{P}',\vec{P}) P_q \left[\phi_A(\vec{p}_{A}) \phi_B(\vec{p}_{B}) \delta^{(3)}(\vec{P}-\vec{P}_{i}) \right] \,,\\
&
{}^{\rm RGM}N_{E}(\vec{P}',\vec{P}_{i}) = \int d\vec{p}_{A'}
d\vec{p}_{B'} d\vec{p}_{A} d\vec{p}_{B} d\vec{P} \phi_{A'}^{\ast}(\vec{p}_{A'}) \times \nonumber \\
&
\times  \phi_{B'}^{\ast}(\vec{p}_{B'})
P_q \left[\phi_A(\vec{p}_{A}) \phi_B(\vec{p}_{B}) \delta^{(3)}(\vec{P}-\vec{P}_{i}) \right] \,,
\end{align}
\end{subequations}
where ${\cal H}$ is the Hamiltonian at quark level.

We will perform a coupled-channels calculation and so the mass and properties of the $T_{QQ'[\bar s\bar s]}$ molecular candidates will be obtained from the $T$-matrix poles, calculated via the Lippmann-Schwinger equation
\begin{align} \label{ec:Tonshell}
T_\beta^{\beta'}(z;p',p) &= V_\beta^{\beta'}(p',p)+\sum_{\beta''}\int dq\,q^2\,
V_{\beta''}^{\beta'}(p',q) \nonumber \\
&
\times \frac{1}{z-E_{\beta''}(q)}T_{\beta}^{\beta''}(z;q,p) \,,
\end{align}
where $\beta$ encodes the required set of quantum numbers that uniquely determine a partial wave in the meson-meson channel, $V_{\beta}^{\beta'}(p',p)$ is the full RGM potential from Eq.~\eqref{eq:fullpot} and $E_{\beta''}(q)$ is the energy for the momentum $q$.


\section{RESULTS}
\label{sec:results}

We analyze the possible existence of heavy meson-meson molecules with minimum quark content $QQ^\prime\bar q\bar q$, with $Q,Q'=\{c,b\}$ and $q=\{n,s\}$, where $n=\{u,d\}$. We focus on the $J^P$ sectors where the molecule can be in a relative $S$-wave, \emph{i.e.} the positive parity $J^P=0^+$, $1^+$ and $2^+$ cases. To be specific, we can have two pseudo-scalar mesons in $0^+$ channel, either pseudoscalar-vector or vector-vector molecules in $1^+$ case, and just vector-vector molecules in $2^+$ sector.


\begin{table}[!t]
\caption{\label{tab:Tcc} Results for the $cc\bar n\bar n$ sector ($T_{cc}$ states) from a $D^{(*)}D^{(*)}$ coupled-channels calculation with total quantum numbers $J^P=0^+$, $1^+$ and $2^+$. The following meson-meson channels are included in the calculation (in parenthesis the threshold's mass in MeV): $D^0D^{*\,+}$ (3875.10), $D^+D^{*\,0}$ (3876.51) and $D^{*\,0}D^{*\,+}$ (4017.11). Masses, widths and binding energies (referred to the nearest threshold $E_B=m_{\rm thres}-m$) are shown in MeV. The Fifth and sixth columns indicates, respectively, the isoscalar probability of the molecule and the nature of the $T$-matrix pole. Note here that these results were previously calculated in Ref.~\cite{Ortega:2022efc}.}
\begin{ruledtabular}
\begin{tabular}{crrrrr}
 $J^P$ & Mass & Width & $E_B$ & ${\cal P}_{I=0}$ & Type\\
\hline
 $0^+$ & $4016.9$ & $0.60$ & $0.200$  & $0.0\%$ & Virtual \\
 $1^+$ & $3874.7$ & $0.08$ & $0.387$ & $81.3\%$ & Bound \\
 $1^+$ & $3876.5$ & $0.36$ & $0.003$ & $57.7\%$ & Bound \\
 $1^+$ & $4014.0$ & $0.00$ & $3.100$ & $98.6\%$ & Virtual \\
\end{tabular}
\end{ruledtabular}
\end{table}

\begin{table}[!t]
\caption{\label{tab:Tbb} Results for the $bb\bar n\bar n$ sector ($T_{bb}$ states) from a $\bar B^{(*)}\bar B^{(*)}$ coupled-channels calculation with total quantum numbers $J^P=0^+$, $1^+$ and $2^+$. The following meson-meson channels are included in the calculation (in parenthesis the threshold's mass in MeV): $\bar B\bar B$ (10559), $\bar B\bar B^{*}$ (10604.15) and $\bar B^{*}B^{*}$ (10649.42). Masses, widths and binding energies (referred to the nearest threshold $E_B=m_{\rm thres}-m$) are shown in MeV. The Fifth and sixth columns indicates, respectively, the isoscalar probability of the molecule and the nature of the $T$-matrix pole. Note here that these results were previously calculated in Ref.~\cite{Ortega:2022efc}.}
\begin{ruledtabular}
\begin{tabular}{crrrrr}
 $J^P$ & Mass & Width & $E_B$ & ${\cal P}_{I=0}$ & Type\\
\hline
$0^+$ & $10545.9$ & $0.0$  & $13.1$ & $0\%$  & Bound\\
$0^+$ & $10672.6$ & $72.0$ & $-23.2$ & $0\%$ & Resonance \\
$1^+$ & $10582.2$ & $0.0$  & $21.9$ & $100\%$  & Bound\\
$1^+$ & $10593.5$ & $0.0$  & $10.5$ & $0\%$  & Bound \\
$2^+$ & $10642.3$ & $0.0$  & $7.1$ & $0\%$ & Bound \\
\end{tabular}
\end{ruledtabular}
\end{table}

\subsection{The $\mathbf{cc\bar n\bar n}$ and $\mathbf{bb\bar n\bar n}$ sectors}

The $cc\bar n\bar n$ sector is relevant by itself due to the recent discovery of the $T_{cc}^+$ signal by the LHCb collaboration~\cite{LHCb:2021vvq, LHCb:2021auc}. This state was thoroughly analyzed by us using the same model employed herein in Ref.~\cite{Ortega:2022efc}; moreover, its $cc\bar n\bar n$ and $bb\bar n\bar n$ partners with $J^P=0^+$, $1^+$ and $2^+$, as well as isospin $0$ and $1$, were also studied. With the aim of giving a self consistent description of the $QQ^\prime\bar n\bar n$ and $QQ^\prime\bar s\bar s$ sectors, the main results of Ref.~\cite{Ortega:2022efc} are collected in Tables~\ref{tab:Tcc} and~\ref{tab:Tbb}. Note herein that the calculation in the $cc\bar n\bar n$ sector was done considering charge basis in order to explore possible isospin breaking effects; this is because the energy difference between $D^0D^{*\,+}$ and $D^+D^{*\,0}$ is roughly $1.4\,\text{MeV}$, larger than the binding energy of the $T_{cc}^+$. This situation is not replicated in the other doubly heavy tetraquark sectors and then they have been calculated taking into account just the isospin basis.

The $T_{cc}^+$ is determined to be a mostly $J^P=1^+$ $D^0D^{*\,+}$ molecule, with modest contributions of $D^+D^{*\,0}$ and $D^{*\,+}D^{*\,0}$ channels. In addition of the $T_{cc}^+$ candidate, a dubbed $T_{cc}^\prime$ state was found below the $D^+D^{*\,0}$ threshold that could provide a small bump in this energy region; however, considering the theoretical uncertainty, its existence is not definitive. Besides, two virtual states below the $D^*D^*$ threshold in the $0^+$ and $1^+$ sectors were distinguished. Concerning the bottom sector, a resonance in the $0^+$ channel was located above the $\bar B^*\bar B^*$ threshold and, furthermore, four bound states were determined: a $0^+$ bound state below the $\bar B\bar B$ threshold, a $2^+$ bound state below the $\bar B^*\bar B^*$ threshold and two $1^+$ $\bar B\bar B^*$ bound states, one with isospin zero, $I=0$, and the other with $I=1$, which are the heavy partners of the $T_{cc}$ and $T_{cc}^\prime$.


\begin{table*}[!t]
\caption{\label{tab:Tbc} Coupled-channels calculation of the $J^P=0^+$, $1^+$ and $2^+$ $bc\bar n \bar n$ sectors ($T_{bc}$ states). The following meson-meson channels are included in the calculation (in parenthesis the threshold's mass in MeV): $\bar BD$ (7146.59), $\bar B^*D$ (7191.95), $\bar BD^*$ (7287.90) and $\bar B^*D^*$ (7333.26). Errors are estimated by varying the strength of the potentials by $\pm10\%$. $J^P$: Pole's quantum numbers; $I$: Isospin of the state; $M_{\text{pole}}$: Pole's mass in MeV; $\Gamma_{\text{pole}}$: Pole's width in MeV; $E_B$: Pole's binding energy (referred to the nearest threshold $E_B=m_{\rm thres}-m$) in MeV; \emph{RS:} Refers to $\bar BD$, $\bar B^*D$, $\bar BD^*$ and $\bar B^*D^*$ Riemann sheets, respectively, with $F$ meaning first, $S$ second and `$-$' closed channel; ${\cal P}$: Channel probabilities in \%; ${\cal BR}$: Branching ratios in \%.}
\begin{ruledtabular}
\begin{tabular}{cccccccccccccc}
$J^P$ & $I$ & $M_{\text{pole}}$ & $\Gamma_{\text{pole}}$ & $E_B$ & RS  & ${\cal P}_{\bar BD}$ & ${\cal P}_{\bar B^*D}$ & ${\cal P}_{\bar BD^*}$ & ${\cal P}_{\bar B^*D^*}$  & ${\cal BR}_{\bar BD}$ & ${\cal BR}_{\bar B^*D}$ & ${\cal BR}_{\bar BD^*}$ & ${\cal BR}_{\bar B^*D^*}$ \\
\hline
$0^+$ & 0  & $7426_{-10}^{+13}$ & $139_{-14}^{+24}$ & $-92_{-13}^{+10}$ & (S,-,-,S)  & $11 \pm 2$  & - & - & $89 \pm 2$   & $14.2_{-0.9}^{+0.4}$  & - & - & $85.9_{-0.6}^{+0.8}$ \\
$0^+$ & 1 & $7146_{-2}^{+1}$ & $0$ & $1_{-1}^{+2}$ & (F,-,-,S)  & $99 \pm 1$  & - & - & $1 \pm 1$   & $0$  & - & - & $0$  \\
$1^+$ & 0 & $7181_{-11}^{+7}$ & $0$ & $11_{-7}^{+11}$ & (-,F,S,S)  & - & $88_{-6}^{+5}$  & $5 \pm 2$  & $7_{-3}^{+4}$ & - & $0$ & $0$  & $0$  \\ 
$2^+$ & 0 & $7465 \pm 2$ & $149_{-13}^{+15}$ & $-132 \pm 2$ & (S,S,S,S)  & $8.9 \pm 0.1$  & $0.25_{-0.00}^{+0.01}$  & $0.44_{-0.06}^{+0.00}$  & $90.5_{-0.2}^{+0.1}$   & $6.0_{-0.5}^{+0.4}$  & $0$  & $0$  & $93.9_{-0.4}^{+0.5}$  \\ 
\end{tabular}
\end{ruledtabular}
\end{table*}

\subsection{The $\mathbf{bc\bar n\bar n}$ sector}

There are no experimental signals related with the dubbed $T_{bc}$ candidates; in fact, they are not even settled by theory since different approaches predict either its absence~\cite{Eichten:2017ffp, Braaten:2020nwp} or its weakly/deeply bound nature~\cite{Lee:2009rt, Karliner:2017qjm, Guo:2021yws, Deng:2018kly, Chen:2013aba, Kim:2022mpa, Dai:2022ulk, Deng:2021gnb, Ke:2022vsi, Li:2012ss, Radhakrishnan:2024ihu, Padmanath:2023rdu, Alexandrou:2023cqg}. The $bc\bar n\bar n$ case is similar to the $cc\bar n\bar n$ and $bb\bar n\bar n$ ones, the main difference is that we do not have identical mesons involved in the first case, so the antisymmetry operator acting on the molecular wave function is given by just ${\cal A}=(1-P_q)$.

Then, we have performed a constituent-quark-model based meson-meson coupled-channels calculation taking into account the $J^P=0^+$, $1^+$ and $2^+$ quantum numbers, in a isospin base, and including the $\bar BD$ (7146.59), $\bar B^*D$ (7191.95), $\bar BD^*$ (7287.90) and $\bar B^*D^*$ (7333.26) thresholds. Our results are collected in Table~\ref{tab:Tbc}. Two wide resonances are found above the $\bar B^*D^*$ threshold, both in the isoscalar sector but with quantum numbers $0^+$ and $2^+$, respectively. Additionally, two bound states are obtained in the isovector $0^+$ and isoscalar $1^+$ sectors. The $(I)J^P=(1)0^+$ bound state is a shallow molecule, slightly below the $\bar BD$ threshold. The $(0)1^+$ state is a $\bar B^*D$ molecule that can be understood as the $T_{bc}$ partner of the $T_{cc}$ and $T_{bb}$ ones.

As the reader can realize, the situation described here is essentially an intermediate case between the ones found for the $cc\bar n\bar n$ and $bb\bar n\bar n$ sectors. Besides, comparing with Lattice-QCD studies, our prediction of a $(I)J^P=(0)1^+$ bound state is in line with the results of Refs.~\cite{Padmanath:2023rdu, Alexandrou:2023cqg, Junnarkar:2018twb}. On the contrary, we do not find an isoscalar $0^+$ $\bar BD$ bound state as in Refs.~\cite{Radhakrishnan:2024ihu, Alexandrou:2023cqg} but an isovector $0^+$ shallow state. Unfortunately, up to our knowledge, the $I=1$ region has not been explored in current Lattice-QCD analyses.


\begin{table*}[!t]
\caption{\label{tab:Tccss} Coupled-channels calculation of the $J^P=0^+$, $1^+$ and $2^+$ $cc\bar s \bar s$ sector ($T_{cc\bar s\bar s}$ states). The following meson-meson channels are included in the calculation (in parenthesis the threshold's mass in MeV): $D_sD_s$ (3936.70), $D_sD_s^*$ (4080.55) and $D_s^*D_s^*$ (4224.40). Errors are estimated by varying the strength of the potentials by $\pm10\%$. $J^P$: Pole's quantum numbers; $M_{\text{pole}}$: Pole's mass in MeV; $\Gamma_{\text{pole}}$: Pole's width in MeV; $E_B$: Pole's binding energy (referred to the nearest threshold $E_B=m_{\rm thres}-m$) in MeV; \emph{RS:} Refers to $D_sD_s$ and $D_s^*D_s^*$ Riemann sheets, respectively, with $F$ meaning first, $S$ second and `$-$' closed channel; ${\cal P}$: Channel probabilities in \%; ${\cal BR}$: Branching ratios in \%.}
\begin{ruledtabular}
\begin{tabular}{ccccccccc}
$J^P$ & $M_{\text{pole}}$ & $\Gamma_{\text{pole}}$ & $E_B$ & RS  & ${\cal P}_{D_sD_s}$ & ${\cal P}_{D_s^*D_s^*}$ & ${\cal BR}_{D_sD_s}$ & ${\cal BR}_{D_s^*D_s^*}$ \\[0.5ex]
\hline
$0^+$ & $4252_{-2}^{+1}$ & $140_{-10}^{+11}$ & $-28_{-1}^{+2}$ & (S,S)  & $1.8_{-0.1}^{+0.2}$  & $98.2_{-0.2}^{+0.1}$   & $10$  & $90$  \\
\end{tabular}
\end{ruledtabular}
\end{table*}

\begin{table*}[!t]
\caption{\label{tab:Tbcss} Coupled-channels calculation of the $J^P=0^+$, $1^+$ and $2^+$ $bc\bar s \bar s$ sector ($T_{bc\bar s\bar s}$ states). The following meson-meson channels are included in the calculation (in parenthesis the threshold's mass in MeV): $\bar B_sD_s$ (7335.23), $\bar B_s^*D_s$ (7383.75), $\bar B_sD_s^*$ (7479.08) and $\bar B_s^*D_s^*$ (7527.60). Errors are estimated by varying the strength of the potentials by $\pm10\%$. $J^P$: Pole's quantum numbers; $M_{\text{pole}}$: Pole's mass in MeV; $\Gamma_{\text{pole}}$: Pole's width in MeV; $E_B$: Pole's binding energy (referred to the nearest threshold $E_B=m_{\rm thres}-m$) in MeV; \emph{RS:} Refers to $\bar B_sD_s$, $\bar B_s^*D_s$, $\bar B_sD_s^*$ and $\bar B_s^*D_s^*$ Riemann sheets, respectively, with $F$ meaning first, $S$ second and `$-$' closed channel; ${\cal P}$: Channel probabilities in \%; ${\cal BR}$: Branching ratios in \%.}
\begin{ruledtabular}
\begin{tabular}{ccccccccccccc}
$J^P$ & $M_{\text{pole}}$ & $\Gamma_{\text{pole}}$ & $E_B$ & RS & ${\cal P}_{\bar B_sD_s}$ & ${\cal P}_{\bar B_s^*D_s}$ & ${\cal P}_{\bar B_sD_s^*}$ & ${\cal P}_{\bar B_s^*D_s^*}$  & ${\cal BR}_{\bar B_sD_s}$ & ${\cal BR}_{\bar B_s^*D_s}$ & ${\cal BR}_{\bar B_sD_s^*}$ & ${\cal BR}_{\bar B_s^*D_s^*}$ \\[0.5ex]
\hline
$0^+$ & $7329.7 \pm 0.1$ & $6.3_{-0.5}^{+0.4}$ & $5.5 \pm 0.1$ & (S,-,-,F)  & $99.76 \pm 0.02$ & - & - & $0.24 \pm 0.02$   & $0$  & - & - & $0$  \\
$1^+$ & $7476 \pm 1$ & $1.6_{-0.0}^{+0.6}$ & $3 \pm 1$ & (-,S,S,F)  & - & $12_{-5}^{+6}$  & $85_{-7}^{+6}$  & $2.7_{-0.7}^{+0.4}$   & - & $100$  & $0$  & $0$  \\
\end{tabular}
\end{ruledtabular}
\end{table*}

\begin{table*}[!t]
\caption{\label{tab:Tbbss} Coupled-channels calculation of the $J^P=0^+$, $1^+$ and $2^+$ $bb\bar s \bar s$ sector ($T_{bb\bar s\bar s}$ states). The following meson-meson channels are included in the calculation (in parenthesis the threshold's mass in MeV): $\bar B_s\bar B_s$ (10733.76) $\bar B_s\bar B_s^*$ (10782.28) and $\bar B_s^*\bar B_s^*$ (10830.80). Errors are estimated by varying the strength of the potentials by $\pm10\%$. $J^P$: Pole's quantum numbers; $M_{\text{pole}}$: Pole's mass in MeV; $\Gamma_{\text{pole}}$: Pole's width in MeV; $E_B$: Pole's binding energy (referred to the nearest threshold $E_B=m_{\rm thres}-m$) in MeV; \emph{RS:} Refers to $\bar B_s\bar B_s^*$ Riemann sheet, with $F$ meaning first and $S$ second; ${\cal P}$: Channel probabilities in \%; ${\cal BR}$: Branching ratios in \%.}
\begin{ruledtabular}
\begin{tabular}{ccccccc}
$J^P$ & $M_{\text{pole}}$ & $\Gamma_{\text{pole}}$ & $E_B$ & RS  & ${\cal P}_{\bar B_s\bar B_s^*}$   & ${\cal BR}_{\bar B_s\bar B_s^*}$  \\[0.5ex]
\hline
$1^+$ & $10781.4_{-0.5}^{+0.4}$ & $0$ & $0.8_{-0.4}^{+0.5}$ & (S)  & $100$    & $0$  \\
\end{tabular}
\end{ruledtabular}
\end{table*}

\subsection{The $\mathbf{cc\bar s\bar s}$, $\mathbf{bc\bar s\bar s}$ and $\mathbf{bb\bar s\bar s}$ sectors}

The worldwide interest in hidden-strange partners of the $T_{cc}^+$ signal has encouraged the Belle collaboration to perform a search for possible $X_{cc\bar s\bar s}$ tetraquark states in the decays of $\Upsilon(1S)$ and $\Upsilon(2S)$, as well as in $e^+e^-$ annhilitaion reactions, into $D_s^+D_s^+$ and $D_s^{*\,+}D_s^{*\,+}$ final states~\cite{Belle:2021kub}. No significant signals have been observed; however, this does not exclude yet the possible existence of a double-heavy, double-strange states. In fact, many theoretical studies have analyzed the $T_{cc\bar s\bar s}$ system finding few candidates, for example, Refs.~\cite{Yang:2020fou, Deng:2021gnb, Ke:2022vsi, Li:2012ss, Dai:2022ulk}.

In our formalism, we can straightforwardly study the $QQ^\prime \bar s\bar s$ system, just replacing the light antiquarks, $\bar n$, by strange ones, $\bar s$. The main difference with respect to the light analogs is that pion exchanges are forbidden between mesons, which could either weaken or strengthen the resulting meson-meson interaction depending on the quantum numbers under scrutiny.

Then, we have performed a meson-meson coupled-channels calculation of the $J^P=0^+$, $1^+$ and $2^+$ $QQ^\prime\bar s\bar s$ systems. In these cases, the isospin can only be zero and the included meson-meson channels are analog to the $QQ^\prime\bar n\bar n$ counterparts, that is to say (in parenthesis the threshold's mass in MeV):
\begin{itemize}
\item For the $T_{cc\bar s\bar s}$ case: $D_sD_s$ (3936.70), $D_sD_s^*$ (4080.55) and $D_s^*D_s^*$ (4224.40).
\item For the $T_{bc\bar s\bar s}$ case: $\bar B_sD_s$ (7335.23), $\bar B_s^*D_s$ (7383.75), $\bar B_sD_s^*$ (7479.08) and $\bar B_s^*D_s^*$ (7527.60).
\item For the $T_{bb\bar s\bar s}$ case: $\bar B_s\bar B_s$ (10733.76), $\bar B_s\bar B_s^*$ (10782.28) and $\bar B_s^*\bar B_s^*$ (10830.80).
\end{itemize}

Our results are shown in Tables~\ref{tab:Tccss},~\ref{tab:Tbcss} and~\ref{tab:Tbbss} for $T_{cc\bar s\bar s}$, $T_{bc\bar s\bar s}$ and $T_{bb\bar s\bar s}$, respectively. The doubly heavy, doubly strange tetraquarks present generally a weaker interaction and thus the number of found poles are less than in the lighter sectors. In the $T_{cc\bar s\bar s}$ case, a resonance with spin-parity $J^P=0^+$ is found, whose structure is mainly $D_s^*D_s^*$ molecule. Additionally, in the $T_{bc\bar s\bar s}$ sector, we find two virtual states in the $0^+$ and $1^+$ channels as $\bar B_sD_s$ and $\bar B_sD_s^*$ molecules, respectively. Finally, in the $T_{bb\bar s\bar s}$ sector, a $1^+$ virtual state is identified as a $\bar B_s\bar B_s^*$ molecule, which can be interpreted as the strange partner of the $T_{bb}$ state.

If one compares our results with those obtained in previous theoretical studies, Refs.~\cite{Deng:2021gnb, Dai:2022ulk} do not find $T_{QQ\bar s\bar s}$ candidates whereas a $0^+$ $\bar B_s^*\bar B_s^*$ molecule is suggested in Ref.~\cite{Ke:2022vsi} and two $bb\bar s\bar s$ candidates with $J^P=1^+$ and $2^+$, respectively, are determined in Ref.~\cite{Li:2012ss}. Therefore, the doubly heavy, doubly strange tetraquark system could be an interesting system to explore experimentally, and their experimental observation could established the underlying dynamics of two heavy mesons with strange content.


\section{SUMMARY}
\label{sec:summary}

The possible existence of molecular candidates with minimum quark content $QQ^\prime\bar q\bar q$, with $q=\{u,d,s\}$ and $Q=\{c,b\}$, have been assessed in a constituent-quark-model based meson-meson coupled-channels approach that correctly described the properties of the $T_{cc}^+$ resonance~\cite{Ortega:2022efc}. 

We have identified several states and computed their properties. In particular, besides the $T_{cc}^+$ resonance below the $D^0D^{*\,+}$ threshold, we find three $T_{cc}$ partners, five $T_{bb}$ states and four $T_{bc}$ candidates. In the $QQ'\bar s\bar s$ sectors, we predict one $T_{cc\bar s\bar s}$, two $T_{bc\bar s\bar s}$ and one $T_{bb\bar s\bar s}$ structures. 

The results presented herein should contribute to the discovery of other double-heavy tetraquark-like signals in the near future, thus advancing our understanding of multiquark states in the heavy quark sector.


\begin{acknowledgments}
This work has been partially funded by
EU Horizon 2020 research and innovation program, STRONG-2020 project, under grant agreement no. 824093;
Ministerio Espa\~nol de Ciencia e Innovaci\'on under grant nos. PID2022-141910NB-I00 and PID2022-140440NB-C22;
and Junta de Andaluc\'ia with contract no. PAIDI FQM-370.
\end{acknowledgments}


\bibliography{PrintTQQqq-biblio}

\end{document}